\documentclass[aps,prl,twocolumn,groupedaddress]{revtex4-1}

\usepackage{graphicx}

\newcommand{\ea}[1]{\begin{eqnarray}#1\end{eqnarray}}

\begin{document}

\title{Metastable criticality and the super Tonks-Girardeau gas }

\author{Mi\l osz Panfil}
\email{m.k.panfil@uva.nl}
\affiliation{Institute for Theoretical Physics, University of Amsterdam, Science Park 904\\
Postbus 94485, 1090 GL Amsterdam, The Netherlands}

\author{Jacopo De Nardis}
\affiliation{Institute for Theoretical Physics, University of Amsterdam, Science Park 904\\
Postbus 94485, 1090 GL Amsterdam, The Netherlands}

\author{Jean-S\'{e}bastien Caux}
\affiliation{Institute for Theoretical Physics, University of Amsterdam, Science Park 904\\
Postbus 94485, 1090 GL Amsterdam, The Netherlands}

\date{\today}

\begin{abstract}
We consider a 1D Bose gas with attractive interactions in an out-of-equilibrium highly excited state containing no bound states. We show that relaxation processes in the gas are suppressed, making the system metastable on long timescales. We compute dynamical correlation functions, revealing the structure of excitations, an enhancement of umklapp correlations and new branches due to intermediate bound states. These features give a clear indication of the attractive regime and can be probed experimentally. We observe that, despite its out-of-equilibrium nature, the system displays critical behaviour: correlation functions are characterised by asymptotic power-law decay described by the Luttinger liquid framework.
\end{abstract}

\maketitle

Quantum many-body systems exhibiting strong correlations are of prime interest from both theoretical and experimental perspectives as they lead to the breakdown of the simple single-particle excitation picture. Indeed, collective modes reign over the low-energy sector, signalling the presence of quantum critical behaviour \cite[and references therein]{GiamarchiBOOK}. An example of such a system is the 1D Bose gas \cite{1963_Lieb_PR_130_1, *1963_Lieb_PR_130_2} which due to recent experimental progress can be realised and probed in and out of equilibrium \cite{2006_Kinoshita_NATURE_440, 2007_Hofferberth_NATURE_449, 2009_Haller_SCIENCE_325, 2012_Trotzky_NATPHYS_8} with tuneable interaction strength \cite{1998_Olshanii_PRL_81}. 

An interesting question is: how strong can the correlation become? Increasing local repulsive interactions ultimately leads to the Tonks-Girardeau gas (TG) \cite{1936_Tonks_PR_50,1960_Girardeau_JMP_1}, which corresponds to infinite interactions and seems to set the limit of correlation strength in the 1D Bose gas. However even stronger correlations are possible if we consider a specific metastable state of a gas with attractive interactions, the super Tonks-Girardeau gas (sTG) \cite{2005_Astrakharchik_PRL_95, 2005_Batchelor_JSTAT_L10001, 2009_Haller_SCIENCE_325, 2010_Chen_PRA_81, 2010_Girardeau_PRA_81, 2010_Muth_PRL_105, 2011_Kormos_PRA_83}.

Attractive point-like interactions drastically modify the dynamics of the gas as they allow the formation of bound states. In fact, the ground state is a macroscopically large molecule \cite{1964_McGuire_JMP_5}, whose dynamical responses can be computed exactly \cite{2007_Calabrese_PRL_98, *2007_Calabrese_JSTAT_P08032}. Instead we consider here a completely out-of-equilibrium state of the system in which there are no bound states and all particles are in a low-lying gaseous state. Contrary to the ground state this state is characterised by a finite energy per particle in the thermodynamic limit. The structure of excitations around it is similar to that found in fermionic systems with the addition of bound states. 

Relaxation in this system thus goes hand-in-hand with the formation of bound states. However, as we will see, local fluctuations are not efficient in bringing particles together, greatly suppressing the rate of formation of bound states. This makes the system sufficiently long-lived to be experimentally observable.

Our starting point is the Lieb-Liniger Hamiltonian (taking $\hbar=2m=1$)
\ea { \label{eq:H_LL}
H_{LL} = -\sum_{i=1}^N \partial_{x_i}^2  + 2c \sum_{j>i}^N \delta\left(x_i - x_j\right),
}
where $N$ is the number of particles and $c$ parametrizes interactions. Throughout the Letter we focus mostly on the attractive regime ($c<0$). We address the correlations explicitly by studying the density-density correlation function defined by its Lehmann representation
\ea { \label{eq:DSF_def}
S(k, \omega) = \frac{2\pi}{L} \sum_{\lambda\in\mathcal{H}} |\langle\lambda|\rho_k|\textrm{sTG}\rangle|^2 \delta\left(\omega - E_{\lambda} + E_{\textrm{sTG}} \right),\;\;\;\;
}
where $\lambda$ labels states in the Hilbert space $\mathcal{H}$, $\rho_k$ is the Fourier transform of the density operator ($\rho_k = \frac{1}{L}\sum_q \Psi_{q+k}^{\dagger}\Psi_q$) and $E_{\lambda}$ is the energy of a state $|\lambda\rangle$. $|\textrm{sTG}\rangle$ is the sTG state and its explicit definition is provided after we discuss the eigenstates. The density-density correlation can be measured using Bragg spectroscopy \cite{2010_Clement_PRL_102}.

The sTG gas can be experimentally realised by employing a confinement induced resonance \cite{1998_Olshanii_PRL_81, 2009_Haller_SCIENCE_325}. Quenching the magnetic field over the resonance changes the sign of the interactions between particles. Starting in the ground state of the TG gas, the interaction parameter is quenched to the other side of the resonance. Quenching to $c=-\infty$ exclusively populates the sTG state. Quenching to a finite but large negative value of $c$ leads to a combination of states with the sTG state being predominant, all other states having an amplitude suppressed by at least a factor $|c|^{-2}$.

The Letter is organised as follows. We start with the description of the eigenstates, first generally for attractive interactions and later specifically around the sTG state. We then discuss the results for the density-density correlation in momentum space and real space, displaying their metastable quantum critical features.

\paragraph{Eigenstates.}

The Lieb-Liniger Hamiltonian (\ref{eq:H_LL}) exhibits an exact solution by the Bethe Ansatz \cite{1963_Lieb_PR_130_1}. The N-particle wave function is given by a linear combination of plane waves 
\ea { \label{eq:EF} 
\Psi(x_1,\dots,x_N) = \prod_{j>k}^N sgn\left(x_j-x_k \right) \sum_{P_N}\mathcal{A}_P e^{i\sum_{i}\lambda_{P_i}x_i}\;\;\;
}
with $\mathcal{A}_P = \left(-1\right)^{[P]} e^{\frac{i}{2}\sum_{j>k}sgn\left(x_j-x_k\right)\phi\left(\lambda_{P_j}-\lambda_{P_k}\right)}$ and the two-particle phase shift $\phi\left(\lambda\right) = 2 \arctan \left(\lambda/c\right)$. By imposing periodic boundary conditions rapidities $\{\lambda_j\}_{j=1}^N$ get quantised and fulfil a set of coupled non-linear equations (Bethe equations)
\ea { \label{eq:BE}
\lambda_j = \frac{2\pi}{L}I_j + \frac{1}{L}\sum_{k=1}^N \phi\left(\lambda_j-\lambda_k\right).
}
The quantum numbers $\{I_j\}_{j=1}^N$ are integers (N odd) or half-odd integers (N even) and completely specify the eigenstate. The Hilbert space of $H_{LL}$ is spanned by different choices of sets $\{I_j\}$. The momentum of the eigenstate is $P_{\lambda} = \sum_{j=1}\lambda_j$, whereas the energy equals $E_{\lambda} = \sum_{j=1}^N \lambda_j^2$.

The structure of the Hilbert space is different depending on the sign of interactions. Characteristic for repulsive interactions is a Pauli-like principle: quantum numbers must be mutually distinct and  all solutions to the Bethe equations are real. Attractive interactions drastically change this by allowing complex solutions \cite{1964_McGuire_JMP_5}. Rapidities then form regular patterns in the complex plane (strings) with exponentially small corrections in the system size which for the system size considered here are completely negligible. They are symmetric around the real axis and their imaginary parts are separated by $|c|$. These complex states are naturally understood as bound states and should be viewed as independent, stable particles. The non-zero imaginary part of rapidities causes exponential decay of the wave function for an increasing separation of two particles within a string with a characteristic distance $|c|^{-1}$ \cite{1964_McGuire_JMP_5}. 

The ground state is formed by a single bound state ($N$-string). All other eigenstates can be constructed by two basic operations, either by breaking them and/or by giving them momentum. In the attractive regime, the Pauli principle holds between strings of the same length. The state classification is then straightforward \cite{2007_Calabrese_JSTAT_P08032}.

\begin{figure}
\includegraphics[scale=0.3, clip, trim = 30 160 0 160]{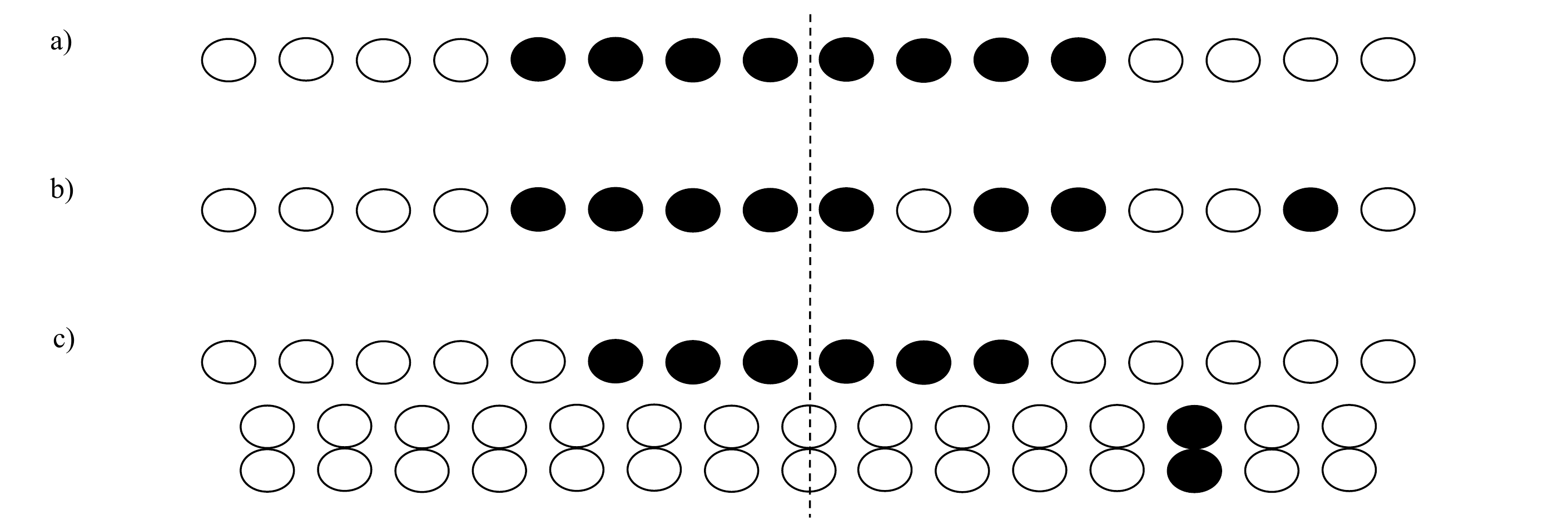}
\caption{\label{fig:States}Pictorial representation of quantum numbers for system with $N=8$ particles and in a) the sTG state, b) an excited state with one particle-hole excitation c) an excited state with one 2-string.}
\end{figure}

\paragraph{sTG gas.}
For a system with N particles we explicitly define the sTG gas by the following choice of quantum numbers
\ea { \label{eq:sTG_QN}
I_j^{sTG} = -\frac{N+1}{2} + j, \;\;\; j=1,\dots, N.
}
The quantum numbers are exactly the same as for the ground state of the repulsive gas of $N$ particles and form a Fermi sea (Fig. \ref{fig:States}) .

In order to compute correlations we first need to classify the excitations around the sTG gas. One class of excitations contains the particle-hole excitations known from the repulsive gas \cite{1963_Lieb_PR_130_2}. We can again distinguish type I and type II modes which can be combined to create multi particle-hole excited states (see Fig. \ref{fig:States}b). These particle-hole excitations lead to a modified version of correlations as compared to the repulsive gas. Attractive interactions induce smooth changes to the particle-hole contribution to the correlation as one tunes the variable $1/c$ through zero.

Other classes of excitations involve bound states. They are directly understood as processes of binding single particles in molecules (see Fig. \ref{fig:States}c) and are ultimately responsible for the instability of the gas. These classes lead to new branches of correlations which we discuss in the next section. Before that however we examine the simplest (and most probable) process: the formation of a $2$-string (bound state with $2$ particles) in order to quantify the timescale for stability of the gas. 

Consider adding a localised impurity potential to the otherwise isolated system. This perturbation connects the sTG state with all the states at the same energy. The timescale can be directly computed using Fermi's golden rule. Assuming that the external potential is weak (we set the magnitude of the potential to be $V = 0.1 \epsilon_F$ ) for a gas of ${}^{133}\textrm{Cs}$ atoms of 1D density $n=10^6 m^{-1}$ with $c/n=-8$ we get $\Gamma_{\textrm{2-str}} = 2\pi \hbar V^2 S(\omega=0^-)/2m \sim 1\, s^{-1}$. $S(\omega=0^-)$ is the density-density correlator summed over all momenta (see Eq. \ref{eq:TSF}) and taken at the slightly negative value of the energy so that only bound states contribute. For larger attractive interactions and weaker perturbations the timescale is much larger. The typical timescale of experiments with cold atomic gases are shorter ($\sim 10^{-3} - 10^{-1} s$) than $\Gamma_{\textrm{2-str}}^{-1}$ and thus the sTG gas is experimentally stable. The rate of formation of bound states is approximately only one order of magnitude larger than the rate of 3-body collisions \cite{2003_Gangardt_PRL_90}. This shows that the sTG gas can be viewed in practice as being as stable as the repulsive gas. 

A physical picture of this statement on stability is as follows. For a $2$-string to form, two particles must be a distance $|c|^{-1}$ apart. This is however highly unlikely since the initial non-local pair correlation function ($S(x)$) exhibits fermionic behaviour: at small distances its value is small. This shows that possible decay channels are not really open (for $c=-\infty$ they are in fact completely closed since the rates from Fermi's golden rule vanish). We further analyse the behaviour of $S(x)$ in the next section. Further evidence for the stability of the sTG gas is provided by numerical computation of the compressibility of the gas which is positive for large attractive interactions \cite{2005_Astrakharchik_PRL_95}.

\paragraph{Density-density correlation function.}
We directly compute the Fourier transform of the density-density correlation function (Eq. \ref{eq:DSF_def}) by adapting the method used previously for the repulsive gas \cite{2006_Caux_PRA_74}. The matrix elements of the density operator (form factors) are known exactly through the methods of Algebraic Bethe Ansatz \cite{1990_Slavnov_TMP_82}. Eq. (\ref{eq:DSF_def}) demonstrates the computational method. The ABACUS algorithm \cite{2009_Caux_JMP_50} was used to recursively explore the Hilbert space looking for states that contribute the most to the correlation. Upon summing contributions we obtained highly accurate results for to the density-density correlation function for $N=128$. The exact f-sum rule identity
\ea {
\int_{-\infty}^{\infty} \omega S(k, \omega) \frac{d\omega}{2\pi} = \frac{N}{L} k^2,
}
gave a quantitative check of the calculations (Tab. \ref{tab:Saturation}). 

\begin{table}
\caption{\label{tab:Saturation}Levels of saturation of the f-sum rule. All computations were performed at unit filling ($N/L = 1$) and for $N=128$ particles.}
\begin{ruledtabular}
\begin{tabular}{c | c  c  c  c}
& $c=-8$ & c=-16 & $c=-64$ & $c=-256$ \\
\hline
$k=k_F$ & 99.7\% & 99.7\% & 99.9\% & 99.9\% \\
$k=2k_F$ & 99.3\% & 99.3\% & 99.9\% & 99.9\% 
\end{tabular}
\end{ruledtabular}
\end{table}

\begin{figure}
\includegraphics[scale=0.42]{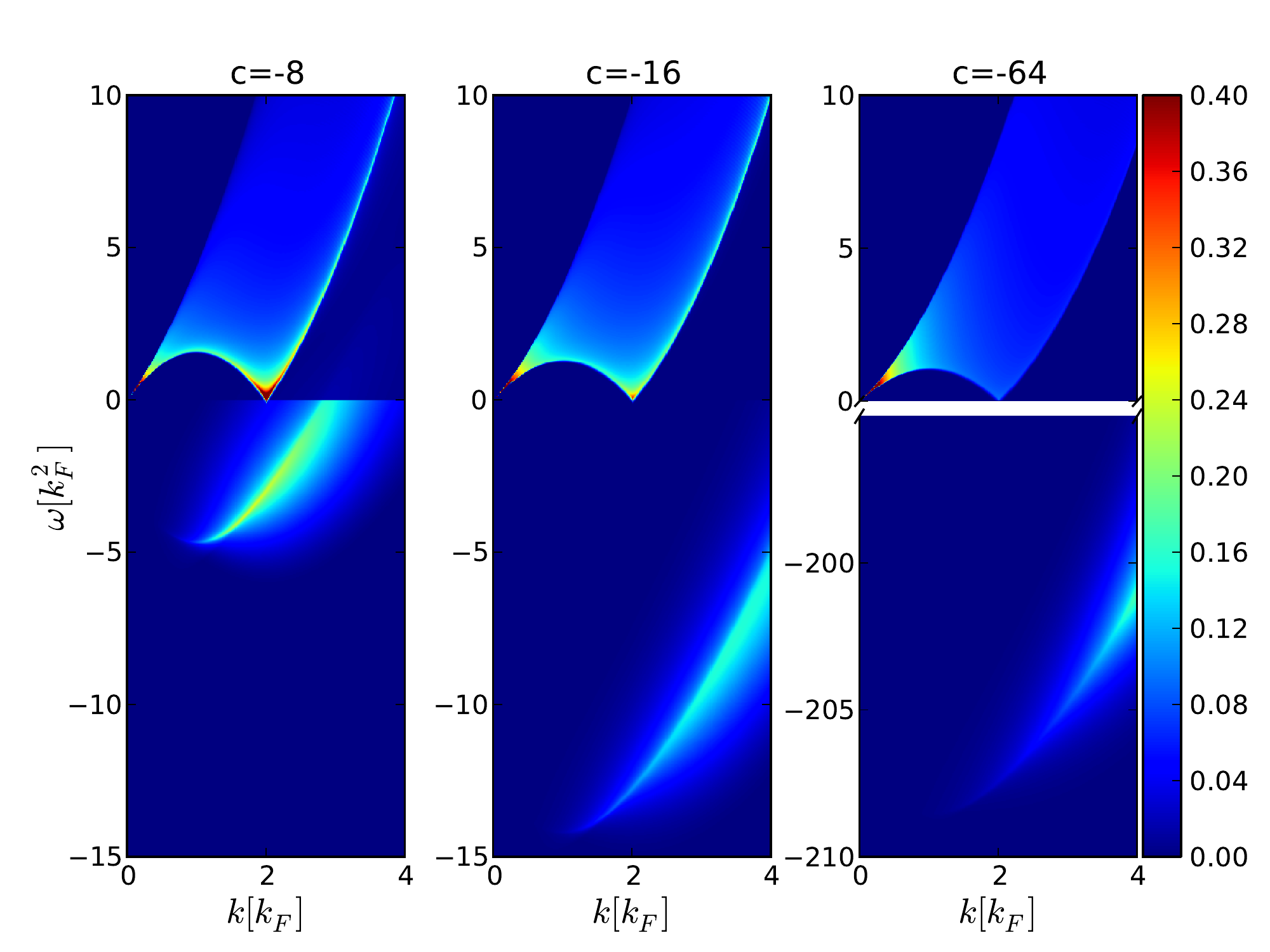}
\caption{\label{fig:DSF}(color online) The density-density correlation in momentum and energy space for $c=-8$, $c=-16$ and $c=-64$. The negative energy parts of the plots were rescaled by factors $10$, $100$ and $2.5 \times 10^5$ respectively to make the string contribution easier to see. The discontinuity in the $c=-8$ plot around $\omega=0$ is an artefact of this rescaling. As the interactions become more attractive the correlation weight spreads uniformly between the lower and upper thresholds of the single particle-hole continuum, just as for the TG gas. At the same time the contribution from bound states is suppressed in value and moves to lower energy.}
\end{figure}

\begin{figure}
\includegraphics[scale=0.4]{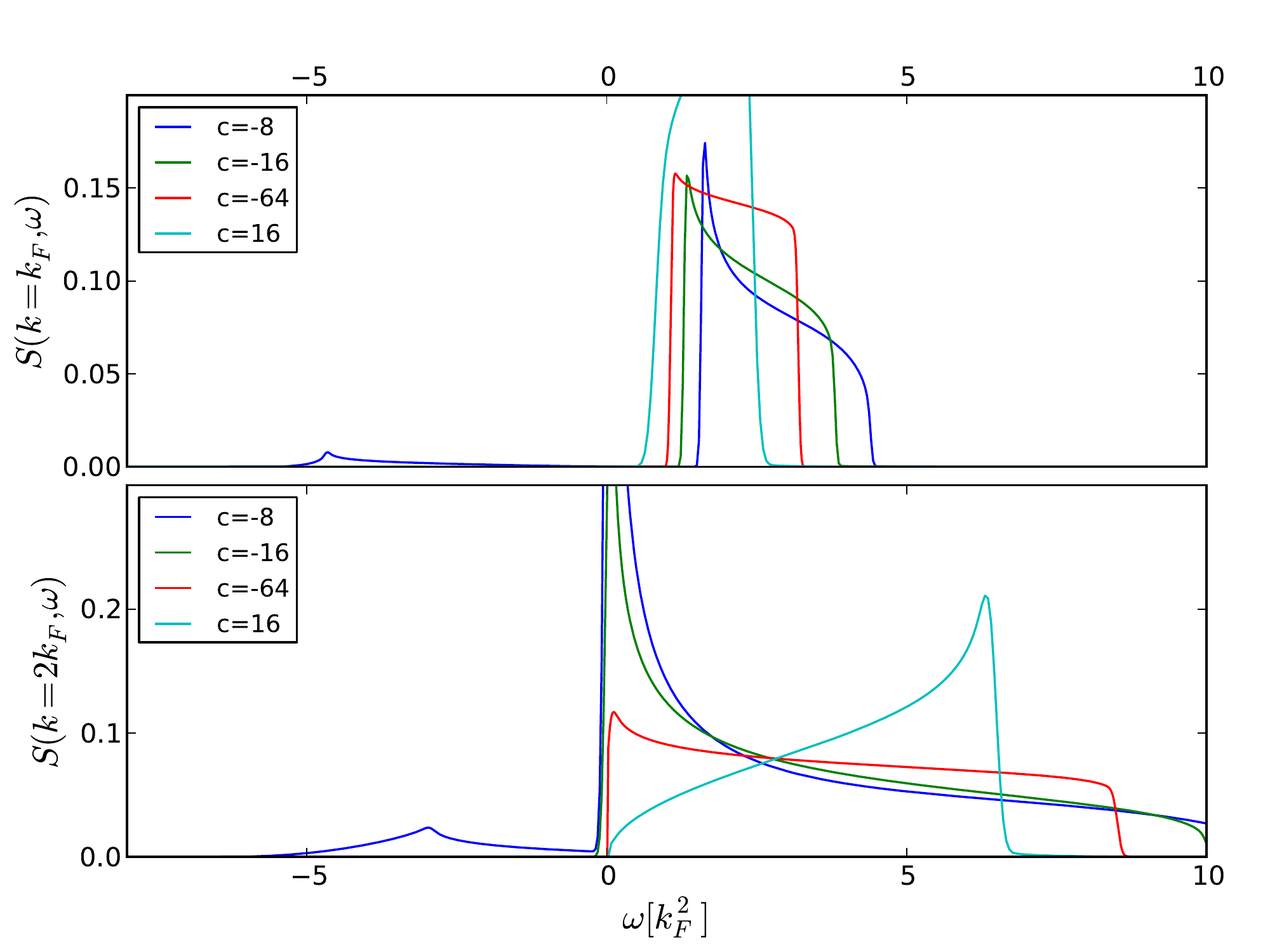}
\caption{\label{fig:fixK}(color online) Fixed momentum cuts through the dynamical structure factor. The shift of the correlation weight towards the lower threshold and the contribution from the bound states for $\omega<0$ are clearly visible. For  comparison the $c=16$ ground state correlation, where the singularity is at the upper threshold, was also plotted.}
\end{figure}

\paragraph{Momentum space.}
The full dynamical correlation is plotted in Fig. \ref{fig:DSF}. As the sTG state takes the form of a Fermi sea, its low-lying type I and II excitations have a linear spectrum and this sector of the sTG gas falls into the Luttinger liquid universality class, with Luttinger parameter $0.5 <K < 1$. The behaviour of the correlation along the edges of support of a single particle-hole excitation agrees in fact with the predictions of non-linear Luttinger theory \cite{2008_Imambekov_PRL_100, *2009_Imambekov_SCIENCE_323}. For values of $K>1$ (repulsive interactions) there is a singularity along the type I mode (upper threshold) and smooth vanishing of correlation along the type II mode (lower threshold). For $0.5<K<1$, inversely to the repulsive case, the correlation is smooth along the particle mode and diverges along the hole mode. This shift of correlation weight towards the lower threshold is exactly what we observe (see Fig. \ref{fig:DSF} and \ref{fig:fixK}).

\begin{figure}
\includegraphics[scale=0.4]{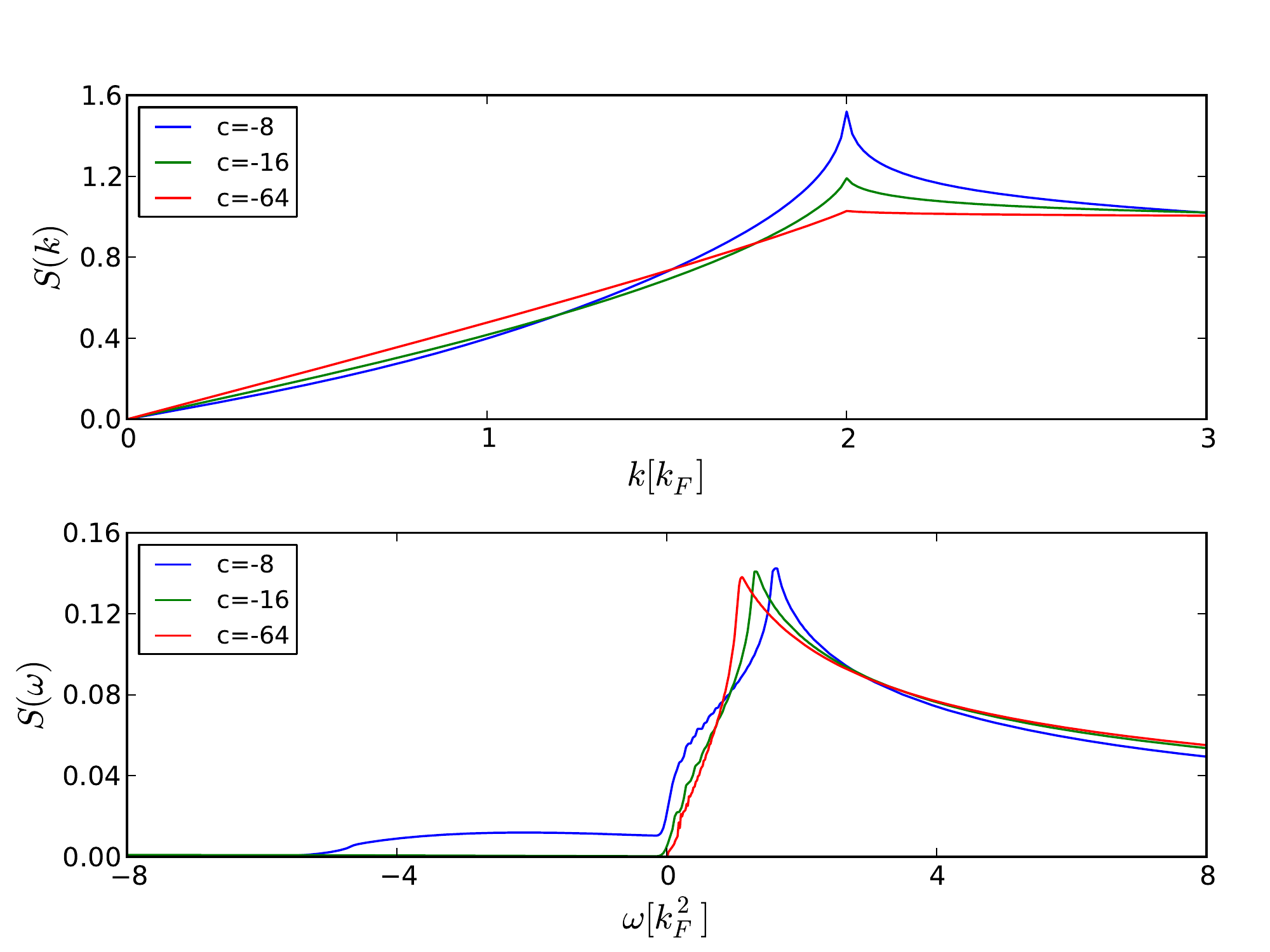}
\caption{\label{fig:SSF_TSF}(color online) \textit{Top:} Static structure factor. Attractive interactions lead to a singularity at $k=2k_F$. \textit{Bottom:} Dynamical autocorrelator. For smaller attractive interactions an extended plateau develops for $\omega < 0$. This clearly indicates the contribution to the density-density correlation from the bound states.}
\end{figure}

Two other interesting quantities namely the static correlator and the dynamical autocorrelator
\ea {
&S(k) = \int_{-\infty}^{\infty} S(k,\omega) \frac{d\omega}{2\pi},\label{eq:SSF}\\
&S(\omega) = \frac{1}{L} \sum_k S(k,\omega),\label{eq:TSF}
}
are plotted in Fig. \ref{fig:SSF_TSF}. The shift of the correlation weight towards the lower threshold leaves a characteristic trademark in the static correlator (Eq. \ref{eq:SSF}, and Fig. \ref{fig:SSF_TSF}) \cite{2005_Astrakharchik_PRL_95}. For repulsive interactions, the static correlator around $k=2k_F$ smoothly approaches its asymptotic value from below ($S(k)\rightarrow 1$). Here at $k=2k_F$ we observe a divergence of the correlator and a power-law tail above the asymptote.

Moreover, for less attractive interactions, the dynamical autocorrelator $S(\omega)$ (Eq. \ref{eq:TSF} and Fig. \ref{fig:SSF_TSF}) develops a plateau on the negative side of $\omega$. This plateau is a clear signature of the attractive gas and of the existence of bound states. Together with the divergence at the lower threshold, these are the two smoking guns to look for experimentally.

\begin{figure}
\includegraphics[scale=0.4]{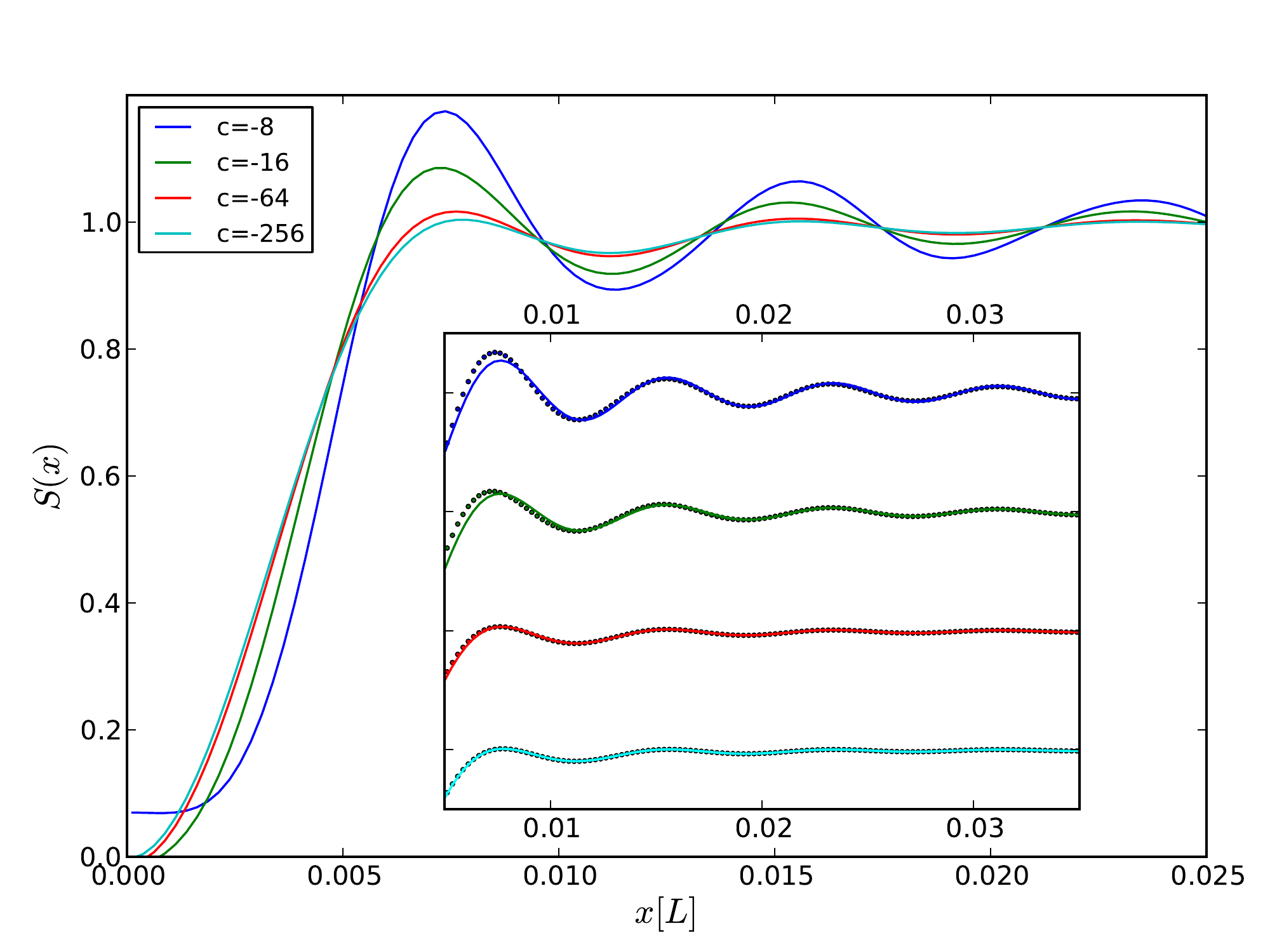}
\caption{\label{fig:SFT}(color online) Distance dependence of the density-density correlation. Despite varying the interaction parameter system is consistently fermionic in behaviour. Smaller attractive interactions increase Friedel oscillations. \textit{Inset:} Correlations immediately approach the asymptotic behaviour predicted by Luttinger liquid theory. All curves converge to 1.}
\end{figure}

\paragraph{Real space.}
We now move on to real space and inspect the Fourier transform of the static correlator
\ea { \label{eq:SFT}
S(x) = \frac{1}{L}\sum_k e^{-ikx} S(k).
}
From the behaviour of $S(x)$ for small $x$ we can infer the effective statistics of the particles (Fig. \ref{fig:SFT}). We see the fermionic-like behaviour which is robust to changes in the interaction \cite{2011_Kormos_PRA_83}. This is similar to the fermionization process in the repulsive 1D Bose gas \cite{1960_Girardeau_JMP_1}.  

At large distances, results can be compared with Luttinger liquid theory which predicts \cite{1981_Haldane_PRL_47, GiamarchiBOOK} the large distance asymptote of correlations as a series
\ea { \label{eq:SFT_LL}
S_{LL}(x) = 1 - \frac{K}{2\left(\pi x\right)^2} + \sum_{m\geq1} A_m\frac{\cos\left(2mk_F x \right)}{x^{2m^2 K}}.
}
This formula is valid in the absence of bound states. These can be captured by treating them as mobile impurities in the gas. The bosonization of the system then leads to a theory similar to that of the ferromagnetic Luttinger liquid \cite{2007_Zvonarev_PRL_99}. However the existence of bound states does not have a strong effect on the spatial correlator giving only subleading corrections which we will investigate in future publications. 

The prefactors $A_m$ in Eq. (\ref{eq:SFT_LL}) are not universal. They depend on the microscopic theory and are connected with the scaling limit of form factors \cite{2011_Shashi_PRB_84, *2012_Shashi_PRB_85}. This in turn allows for the exact computation of prefactors in integrable theories where form factors are explicitly known. Fig. \ref{fig:SFT} depicts the comparison between the correlations computed for $N=128$ and the asymptotics (including prefactors) from \cite{2012_Shashi_PRB_85}. The accurate match is yet another illustration of the applicability of the theory of Luttinger liquids in this out-of-equilibrium context.

\paragraph{Conclusions and outlook.}
In this paper we considered the metastable super Tonks-Girardeau gas and its correlations. We showed that attractive interactions modify the dynamical responses of the system in a clear, experimentally observable way. The system exhibits stronger collective behaviour with the majority of the density-density correlation carried by the type II mode. Attractive interactions strengthen the umklapp excitations leading to an extended region of high correlation around $2k_F$, which ultimately causes the divergence of the static correlator at $k=2k_F$. On top of this there are bound states with an extended region of correlation for $\omega<0$. These features provide experimentally clear signatures of the super Tonks-Girardeau gas, which should be experimentally accessible with current methods.

We also showed that, despite attractive interactions and metastability, the super Tonks-Girardeau gas still displays the standard features of a quantum critical liquid. The system has a sector of excitations which falls into the Luttinger liquid universality class and the leading long-distance asymptotes of correlations agree with the predictions of Luttinger liquid theory. By treating bound states as impurities, a theory similar to the ferromagnetic Luttinger liquid can be developed, allowing to study the dynamics of bound states in more detail. Moreover, initial metastable quantum critical states can exist in other systems (e.g. multicomponent bosons, fermions). We will investigate these issues in future publications.

\begin{acknowledgments}
We acknowledge useful discussions with H.-C. N\"agerl and V. Gritsev. We thank SARA Computing and Networking Services for access to the Lisa Compute Cluster. We gratefully acknowledge support from the Foundation for Fundamental Research on Matter (FOM) and from the Netherlands Organisation for Scientific Research (NWO).
\end{acknowledgments}

\bibliography{/Users/milosz/Documents/BIBTEX_LIBRARY_JSCaux_PAPERS,/Users/milosz/Documents/BIBTEX_LIBRARY_JSCaux_BOOKS,/Users/milosz/Documents/BIBTEX_LIBRARY_JSCaux_OWNPAPERS}

\end{document}